\documentclass[runningheads]{llncs}
\usepackage[T1]{fontenc}
\usepackage{graphicx}
\usepackage{hyperref}
\usepackage{color}

\usepackage{cite}
\usepackage{todonotes}
\usepackage{xurl}

\begin{document}

\title{Age Verification in the Web - Holy Grail to Control Access to Restricted Content}

\titlerunning{Age Verification in the Web - Access Control to Restricted Content}

\author{Wojciech Wodo\inst{1}
\and
Maksymilian Gorski\inst{2}
\and
Lucjan Hanzlik\inst{2}
}

\authorrunning{W. Wodo et al.}

\institute{Wroclaw University of Science and Technology\\
Wybrzeze Wyspianskiego 27, 50-370 Wroclaw, Poland\\
\email{wojciech.wodo@pwr.edu.pl}\\
\and
CISPA Helmholtz Center for Information Security, Germany\\
\email{maksymilian.gorski@cispa.de, hanzlik@cispa.de}}

\maketitle            

\begin{abstract}
Age verification before accessing restricted content is critical to protecting minors from exposure to harmful material such as pornography, gambling, violence, hateful speech, and substance purchases like alcohol and tobacco. Currently, the absence of reliable age-checking mechanisms allows children extensive access to such adult content, posing significant risks to their worldview and mental development. While regulatory efforts like the European Union’s Digital Services Act promote using Digital Wallets or Age Verification Apps, relying solely on government-based solutions raises concerns about data sensitivity and privacy risks. Effective age verification must therefore be trustworthy, user-friendly, privacy-preserving, and offer flexible assurance levels. 
We analyze currently implemented (UK or Australia) and proposed (UE) solutions from different angles, pointing out the weaknesses and threats, and come up with an alternative.
Our proposal addresses these challenges by leveraging open standards - such as Privacy Pass and Privacy Access Tokens - and cryptographic techniques to enable secure, privacy-conscious age verification without requiring specialized software installation. This approach empowers users to select trusted providers from multiple options, reducing the risk of data breaches and ensuring a safer digital environment for minors.

\end{abstract}

\section{Introduction}
\label{sec:intro}
The issue of age verification is getting a lot of publicity these days. The rationale behind it lies in the restricted content that is easily accessible to minors or unauthorized individuals. Contemporary safeguards and verification mechanisms are imperfect in this regard. European Union legislation is attempting to address the issue, and for now, there is a significant belief in the European Union identity. However, it will be several years before these EUDI wallets will be implemented and widely adopted. At this time, there is a strong need for a solution that ensures user trust, does not require the installation of dedicated software for age verification purposes, provides a certain level of credibility, and does not require conducting the verification process each time the resource is accessed. 

In this case, there are seven main methods for age verification listed in the rules, including photo ID checks, face scanning, age estimation tools, credit card checks, digital ID Voting by the parent of a young person, using AI to guess a user age based on the data, and the last one - which is really important - relying on a third party that has already checked the user age. We believe this is one of the best possible approaches.
It is worth mentioning that in Poland, Piotr Konieczny, a cybersecurity expert from Niebezpiecznik.pl, presented his comments on the UK case\footnote{\url{https://www.youtube.com/watch?v=5nJMvCB7JEA}}. However, he demonstrated a case of wishful thinking, which is right in principle but completely infeasible in practice. What Konieczny proposes is a liability shift, where parents should be entirely responsible for their children's access to restricted content. They should be educated and provide appropriate guidance for their children.

In the European Union, there is a belief that EUDI wallets would solve long-standing problems, but this is not actually true. A special application for age verification is currently under development and may be perceived as a mini version of the digital wallet dedicated solely to this purpose. However, we must consider that governmental or quasi-governmental products and services may not evoke immediate trust among the people. According to the saying, “Don't put all eggs into one basket,” we need to diversify and mitigate risk. And that's why we cannot put full trust in any single solution, which, in this case, might be the EUDI wallet. Moreover, this lack of trust in the government is pretty justified. Let's examine the GDPR violation under EU Regulation 2019/1157 of June 20, 2019, on improving the security features of Union citizens' identity cards and residence documents.

In this case, it concerned unlawful processing of personal data, including biometric data – facial scans and fingerprints collected in connection with the issuance of a new identity document. That was the purpose of the processing, but individual authorities did not delete this data after the document was issued. Instead, they retained it in databases, which were then shared with other institutions. In fact, they processed it inconsistently with the purpose for which the data were collected from the data subjects.

According to the Court of Justice of the European Union, the collection and storage of fingerprints and facial images as biometric data requires special protection. It must be limited solely to the temporary storage required to insert the data into the issued ID card, with a guarantee that it will be deleted immediately once this purpose has been fulfilled. That wasn't the case, e.g., in Poland.

Considering the above, we propose a new solution that fulfills user needs. We propose a solution that utilizes existing open components and standards, thereby guaranteeing transparency. Cryptographically guaranteed anonymity and binding to the device/browser will be achieved by the use of Privacy Pass and Apple Private Access Tokens.

The system offers easy scalability; the new entities for accepting and issuing the tokens can be easily whitelisted. There will be an attestation performed by the third party; it could be like a CAPTCHA challenge - from which the Privacy Pass solution originates - but in that case, it is replaced with a Know Your Customer (KYC) process.

If the person meets the attestation requirements, they will be granted a portion of cryptographic tokens that can be spent anonymously on websites that require age verification or attestation, until the pool is exhausted. The user must then be reattested.

The remaining question is: who should be the third-party providers? Who is experienced enough to provide this kind of service? And here, eIDAS 2.0\cite{european_parliament_and_of_the_council_regulation_2024} comes to the rescue: according to the regulation, there is a new kind of qualified and non-qualified trust service provider - issuing and validating electronic attribute certificates. Within these new players, we can definitely find appropriate attesters. These attesters are ID-proofing companies with extensive experience in providing electronic KYC solutions across various scenarios. They are also obliged by ETSI 119 461\cite{etsi_ts_119_461_electronic_2025} to demonstrate countermeasures against presentation and injection attacks and to provide appropriate certifications of their solutions, which makes them a perfect fit for this role.

\subsection{Motivation and Contribution}
The proliferation of internet resources has introduced unprecedented risks to minors, who encounter a digital landscape replete with unregulated content that can be created and disseminated by anyone with minimal accountability. Unsupervised exposure to pornography, violence, hate speech, and promotions of substances like alcohol and tobacco profoundly impacts children's development, contributing to psychological disorders, developmental delays, suicides, and other adverse outcomes. Compounding this vulnerability, many parents lack the technical savvy or awareness to monitor their children's online activities effectively.

To safeguard minors—the most defenseless members of society—access to adult-oriented content must be rigorously monitored and controlled. Responsibility cannot rest solely with families; providers of restricted content and goods bear a shared duty, necessitating robust legal frameworks at national and international levels. This paper serves as a whistleblowing intervention, urgently drawing public and policy attention to the imperative of not only enacting legislation but also enforcing practical measures—such as age verification technologies and compliance standards—to ensure meaningful implementation and achievement of protective goals.

This paper advances the discourse on protecting minors from harmful online content by systematically analyzing three prominent legislative trials in age verification: the United Kingdom's Online Safety Act, Australia's Social Media Minimum Age Bill, and the European Age Verification Solution. We critically evaluate their respective strengths—such as regulatory enforcement and platform accountability—and limitations, including implementation challenges, privacy risks, and evasion tactics.

Building on this analysis, we propose a novel, privacy-preserving age verification framework leveraging established open standards like Privacy Pass and Apple Privacy Access Tokens, which operate without requiring dedicated software on end-user devices. In alignment with eIDAS 2, we advocate engaging Trust Services Providers—experienced in Know Your Customer (KYC) processes—as key stakeholders to ensure scalability and compliance.

Furthermore, we emphasize conducting comprehensive risk analyses to diversify verification methods, addressing over-reliance on single approaches. We highlight diminished public trust in government-mandated solutions, citing the Court of Justice of the European Union's rulings on GDPR violations under Regulation 2019/1157, which underscore surveillance and privacy concerns. Finally, we argue for establishing minimum assurance levels for verification methods to balance efficacy with fundamental rights.

\subsection{Structure}
The paper is structured as follows. Section~\ref{sec:intro} introduces the reader to the topic of the paper, considering its security, compliance, legal, and usability aspects. Then, in Section~\ref{sec:whyAgeVerifIsImportant}, we explain the importance of the \textit{age verification} issue in light of contemporary uncontrolled access to harmful content on the web, presenting statistical data and preliminary unsuccessful attempts to address this challenge. Section~\ref{sec:solution} presents the concept of our solution proposal, including the architecture, actors, data flow, and building components, along with their properties and limitations. In Section~\ref{sec:conclusion}, we conclude our work, pointing out both advantages and drawbacks, and discussing potential enhancements for the future.

\section{Why Age Verification is so important?}
\label{sec:whyAgeVerifIsImportant}

In this chapter, we point out why the age verification issue is so important for the sake of our children and what might be the impact of doing it properly or not. We also discuss pros and cons of already implemented or prepared solutions (UK, Australia, and EU) in this regard.

\subsection{Rationale and impact}
Minors' access to age-restricted content websites is significant and widespread despite restrictions. Research shows almost one in ten young people (9\%) have encountered harmful age-restricted content such as pornography or alcohol online, with many children as young as eight having viewed or purchased such content. Additionally, about half of minors report encountering some form of safety measure meant to restrict access, but these are often easily bypassed by self-declaration of age or other weak verification methods. About 11\% of young users encounter such content weekly, and 7\% see it daily when online\footnote{\url{https://verifymy.io/blog/young-people-encounter-harmful-age-restricted-and-illegal-content-within-minutes-of-going-online/}}. Also, studies have found that 87.71\% of under-18 desktop (no mobile) traffic to pornographic sites is directed to the top 200 sites (ranked according to UK traffic in the month of August 2020) allowing user interaction or sharing, indicating heavy usage by minors on popular adult websites\cite{british_board_of_film_classification_bbfc_bbfc_2021}.

Moreover, many popular social media platforms allow children to set up accounts by simply lying about their age, bypassing age verification, exposing them to inappropriate content\footnote{\url{https://www.sfi.ie/research-news/news/children-social-media/}}. Almost a third of minors aged 8 to 17 have adult profiles on social media. VPNs and prepaid cards are commonly used by teens to circumvent age restrictions online, increasing exposure to risks such as malware and interaction with predators. Recent studies indicate up to 90\% of teens have viewed pornography online, with about 10\% admitting to daily use, reflecting high exposure facilitated by easy internet access and minimal parental awareness.
These statistics reflect the challenges in effectively restricting minors' access to age-restricted content online and the need for better safeguards and education to mitigate exposure to harmful materials.

Efforts to curb access have recently intensified, especially in the EU and UK, where children now must prove their age through secure methods such as facial scans, photo IDs, or credit card checks to access content categorized as harmful, pornography, self-harm, and hate speech. Over a thousand platforms have complied with these measures, though exposure among very young children remains a concern \footnote{\url{https://commission.europa.eu/news-and-media/news/minimising-risks-children-and-young-people-face-online-2025-07-14\_en}}.

The European Commission released its age verification blueprint in July 2025\footnote{\url{https://digital-strategy.ec.europa.eu/en/news/commission-makes-available-age-verification-blueprint}} and an enhanced second version of it in October 2025 to help online platforms implement a robust, user-friendly, and privacy-preserving age verification method across the EU\footnote{\url{https://digital-strategy.ec.europa.eu/en/news/commission-releases-enhanced-second-version-age-verification-blueprint}}. This blueprint allows users to verify their age to access restricted content (e.g., the adult content) without revealing their personal data beyond proof of being over 18. The newer version expands onboarding options to include passports and identity cards alongside eIDs, and offers support for the Digital Credentials API to enable more user-friendly proof presentation.

The blueprint is built on technical specifications aligned with upcoming EU Digital Identity Wallets and is available as an open-source mobile application, customizable by Member States for national contexts. It can be standalone or integrated into digital identity wallets, with the potential to expand to other age restrictions and use cases, such as alcohol purchases.

The solution emphasizes privacy by ensuring that the user's identity is checked only once during proof issuance. The proof itself contains no personally identifiable information and is presented anonymously to service providers. Proofs are issued in one-time-use batches to prevent tracking. Pilot testing is underway in Denmark, France, Greece, Italy, and Spain, involving stakeholders such as platforms and end users. The integration of zero-knowledge proof technology is planned to further enhance privacy. Member States can adapt the app’s branding and language, with the first customized apps expected by early 2026.

This initiative supports the EU Digital Services Act's protection of minors online and sets a reference standard for device-based age verification across the EU, also serving as groundwork for broader age-appropriate service deployment with the European Digital Identity Wallet rollout planned by the end of 2026.

\subsection{Attempts to solve the issue}

In this section, we present three trials of addressing the challenge of age verification for accessing the online content at the federal level. We refer to the Online Safety Act in the United Kingdom, The Social Media Minimum Age Bill in Australia, and the European Age Verification Solution.

\subsubsection{Online Safety Act 2023 in the United Kingdom:}
The UK's age verification system under the Online Safety Act 2023\cite{osa2023} enforces "highly effective" checks from July 25, 2025, targeting platforms with harmful content like pornography, self-harm promotion, or violence, requiring users to prove their age via facial estimation, photo ID, credit cards, or biometrics\footnote{\url{https://shuftipro.com/blog/uk-age-verification-laws-july-2025-compliance-guide/}}.

Users bypassed restricted sites within seconds of enforcement using VPNs to spoof locations and access non-compliant international mirrors, evading UK-specific checks. Selfies with makeup or filters fooled facial age estimation on some platforms, while proxy ID uploads from adults allowed minors access before behavioral flags triggered re-verification\footnote{\url{https://cybernews.com/uk/how-to-use-vpn/get-around-uk-age-verification-requirements/}}.

Early tests revealed weak circumvention resistance, with hackers exploiting unpatched APIs or low-challenge ages (e.g., set at 20-25) where <1\% of 13-17s slipped through via manipulated documents. Privacy concerns amplify risks, as data minimization fails under mass scans, enabling attacks like credential stuffing on verification providers\footnote{\url{https://www.theguardian.com/technology/2025/jul/24/what-are-the-new-uk-online-safety-rules-and-how-will-they-be-enforced}}.

Online Safety Act mandates age verification (precise age confirmation) or age estimation (probabilistic assessment), or both, applied proportionally based on children's risk assessments under Sections 11-12. Methods must be technically accurate, robust against evasion (e.g., <1\% child access rate), reliable across demographics, equitable (no bias), and compliant with UK GDPR/ICO data protection via minimization and no unnecessary storage\footnote{\url{https://www.onlinesafetyact.net/analysis/the-online-safety-act-childrens-duties-age-verification-and-content-moderation-on-user-to-user-services/}}.

Ofcom guidance lists non-exhaustive options platforms select per service risk\footnote{\url{https://www.ofcom.org.uk/online-safety/illegal-and-harmful-content/quick-guide-to-childrens-safety-codes}}:

\begin{itemize}
    \item Facial age estimation: Live selfies/video analyzed by AI for probable age (e.g., X's AI, Snapchat/Discord scans), processed locally without storing biometrics.
    \item Photo ID matching: Passport/driving license uploaded and matched to selfie with liveness detection.
    \item Credit card/bank detail checks: Transaction confirmation via adult-linked cards or open banking APIs signaling the user's majority.
    \item Mobile network operator (MNO) data: Carrier-confirmed adult status from account age or billing.
    \item Digital identity wallets/services: Third-party proofs (e.g., EU DIDs) yielding yes/no adult signals without data sharing.
    \item Email-based estimation: Behavioral/derived age from verified domains, as a supplement.
\end{itemize}

We note a problem with this implementation stemming from the wide range of acceptable age verification methods. Individual methods do not have the same level of trustworthiness due to differences in potential attack vectors and the ease with which a given solution can be bypassed or cheated.

In the case of automatic identity verification systems, including age and vitality estimation from video streams, current solutions are vulnerable to advanced face reconstruction and animation techniques, such as 3D Gaussian Splatting\cite{kerbl_3d_2023} and Gaussian Avatars\cite{qian_gaussianavatars_2024}. Thanks to their photorealism and real-time control of an artificial avatar, an attacker can easily respond to liveliness challenges posed during the process and naturally move individual parts of the avatar's head. Similar attacks were successfully performed by children against a deployed system (e.g., by using avatars from the video game Death Stranding).

\subsubsection{The Social Media Minimum Age Bill in Australia:}
Australia's Social Media Minimum Age Bill, enacted via the Online Safety Amendment (Social Media Minimum Age) Act 2024\cite{SMMinAgeBIll}, mandates that age-restricted social media platforms prevent Australian children under 16 from holding accounts, with enforcement beginning December 10, 2025. Platforms must implement "reasonable steps" for age verification, facing fines up to AU\$50 million for systemic non-compliance, including privacy breaches during verification\footnote{\url{https://www.infrastructure.gov.au/media-communications/internet/online-safety/social-media-minimum-age}}.

Public support for the under-16 ban remains strong at around 77-88\%\footnote{\url{https://srcentre.com.au/wp-content/uploads/2025/07/20250729_SRC_Media-release_Youth_Social-Media_FINAL.pdf}}, driven by parental concerns over youth mental health and online harms, with bipartisan backing and high approval for stricter platform penalties. However, challenges such as enforcement feasibility and potential circumvention via unregulated platforms temper enthusiasm, as noted in ongoing surveys evaluating the impacts on children aged 10-16\footnote{\url{https://srcentre.com.au/aii/project/social-media-minimum-age-survey/}}.

Experts highlight privacy risks from age assurance technologies requiring biometric data or ID sharing, with widespread distrust in platforms' data handling despite mandated protections like data minimization. Tech groups like NetChoice criticize the ban as censorship that undermines digital literacy, while trials show some vendors achieving over 90\% accuracy at age boundaries, though self-declaration methods fail robustness tests\footnote{\url{https://www.scimex.org/newsfeed/expert-reaction-australias-social-media-ban-passes-the-senate}}.

The framework emphasizes flexible, privacy-centered verification - avoiding mandatory government ID - with ongoing monitoring of existing accounts and appeal processes, but lacks precise numerical thresholds, relying on eSafety Commissioner guidelines. Lawsuits filed by Reddit and teens question constitutional validity over free speech and privacy, signal potential vulnerabilities amid rapid tech evolution\footnote{\url{https://www.npr.org/2025/12/12/g-s1-101828/reddit-challenges-australias-social-media-law}}.

Australia's social media minimum age system has faced early bypasses through facial age estimation flaws, proxy verification, and other workarounds, as reported in initial enforcement days\footnote{\url{https://www.abc.net.au/news/2025-12-10/social-media-ban-day-one-teen-access/106126706}}.

Reported bypassing techniques include \textit{Makeup and Appearance Alteration} (e.g. applying fake lashes and makeup), \textit{Proxy Facial Scans} (teens use older siblings, parents, or guardians to perform facial scans during verification), \textit{VPNs} and \textit{Fake IDs}, where children employ VPNs to mask locations and fake identification or outdated birthdates from account creation to evade checks.

In the described case of implementation of age verification solution, similarly to the United Kingdom, too many solutions with a low level of assurance are allowed. The regulator allows methods that also differ significantly in terms of potential attack vectors, including spoofing methods.
This means that an attacker can always reduce the security of the whole solution to the security and vulnerability of the weakest method.

\begin{figure*}[!htp]
\includegraphics[width=\textwidth]{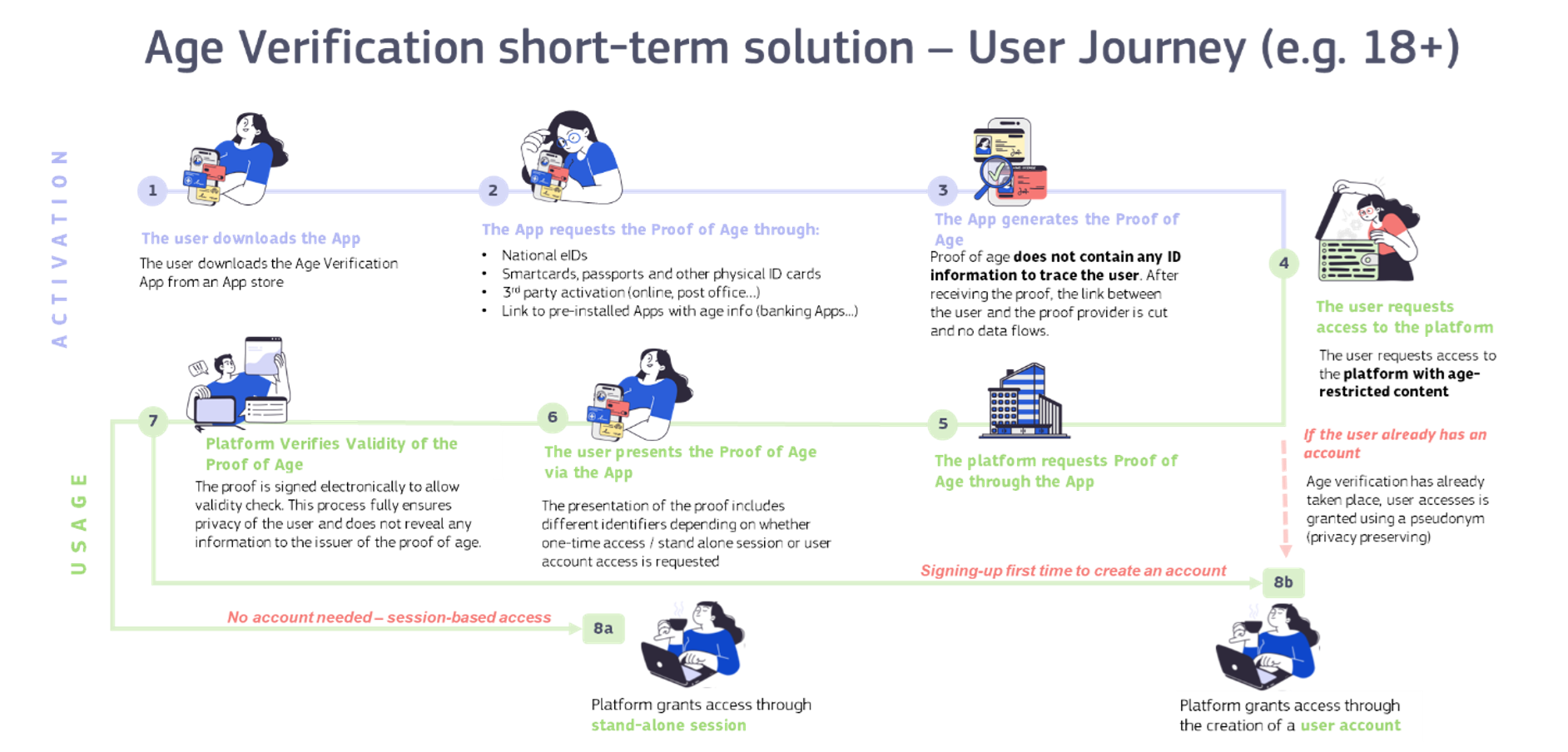}
\caption{UE Age Verification App - Journey of a Proof of Age Attestation User\cite{UEAgeVerificationApp}} \label{fig:uuAppUserJourney}
\end{figure*}

\subsubsection{UE Approach to Age Verification:} 
The European Commission is developing a harmonised, EU-wide age verification system to help online platforms verify users’ age while preserving privacy and user-friendliness\footnote{\url{https://digital-strategy.ec.europa.eu/en/policies/eu-age-verification}}. This initiative supports compliance with the Digital Services Act, starting with verification that users are over 18 to access adult-restricted content like pornography, gambling, and alcohol sales.

On July 14, 2025, the Commission released an age verification blueprint allowing users to confirm they are over 18, without sharing their personal data. This privacy-focused and user-friendly solution is fully compatible with future EU Digital Identity Wallets and adaptable for other age thresholds, such as 13+.

A second blueprint, published on October 10, 2025, adds onboarding via passports and ID cards and supports the Digital Credentials API. Known as the ‘mini wallet,’ this software aligns with the technical standards of European Digital Identity Wallets, planned for launch by the end of 2026.

The solution is now entering a pilot phase involving Member States, online platforms, users, and software providers. It is customizable (e.g., language localization) but maintains mandatory privacy features. The technical specs, source code, and a beta version are openly published\footnote{https://ageverification.dev/Setup/}, supported by a contract awarded to T-Scy (Scytales and T-Systems).

\subsection{Guidelines on the protection of minors}
The guidelines on the protection of minors\footnote{\url{https://ec.europa.eu/commission/presscorner/detail/en/ip\_25\_1820}} ensure children enjoy high levels of privacy, safety and security on online platforms. This follows an inclusive and extensive consultation period, including with young people.
Among other things, the guidelines provide recommendations to address:

\begin{description}
    \item[Addictive design:] Minors are particularly vulnerable to practices that can stimulate addictive behaviour. The guidelines suggest reducing minors' exposure to such practices, and disabling features that promote the excessive use of online services, like "streaks" and "read receipts" on messages.
    \item[Cyberbullying:] The guidelines recommend empowering minors to block or mute users, ensuring they cannot be added to groups without their explicit consent. They also recommend prohibiting accounts from downloading or taking screenshots of content posted by minors to prevent the unwanted distribution of sexualised or intimate content.
    \item[Harmful content:] Some recommender systems put children in harmful situations. The guidelines give young users more control over what they see, calling on platforms to prioritise explicit feedback from users, rather than relying on monitoring their browsing behaviour. If a young user indicates they do not want to see a certain type of content, it should not be recommended again.
    \item[Unwanted contact from strangers:] the guidelines recommend that platforms set minors' accounts that are private by default - that is, not visible to users that are not on their friends' list - to minimise the risk that they are contacted by strangers online.
\end{description}

The guidelines adopt a risk-based approach, like the Digital Services Act, recognising that online platforms may pose different types of risks to minors, depending on their nature, size, purpose and user base. Platforms should make sure that the measures they take are appropriate and do not disproportionately or unduly restrict children's rights.

The guidelines also recommend the use of effective age assurance methods provided that they are accurate, reliable, robust, non-intrusive, and non-discriminatory. In particular, the guidelines recommend age verification methods to restrict access to adult content such as pornography and gambling, or when national rules set a minimum age to access certain services such as defined categories of online social media services.

\subsection{European Age Verification Solution: Age Verification App}

The European Age Verification Solution\footnote{\url{https://ageverification.dev/}} is a harmonized, EU-wide framework developed by the European Commission to enable privacy-preserving age verification across all Member States. It leverages the European Digital Identity Wallet infrastructure to allow individuals to prove their eligibility for age-restricted online services without revealing unnecessary personal information. This solution supports compliance with Article 28 of the Digital Services Act, aiming to help online service providers meet legal requirements for age verification while safeguarding user privacy.

The solution architecture is modular, consisting primarily of four core components: an Issuer Service that issues Proof of Age Attestations, an Age Verification App that securely stores attestations on the user's device, a Verifier Service that validates these attestations for service providers, and a Trusted List to ensure authorized issuers across Member States. The system aims to enable ease of integration and interoperability while preserving security and privacy through open standards. Likewise, the Proof of Age credential lifecycle is designed to be as simple as possible to make it accessible and understandable even to non-technical end users. We provide a suggested "User Journey" through the attestation process in Fig.~\ref{fig:uuAppUserJourney}. The integration of more advanced privacy-enhancing technologies, such as Zero-Knowledge Proofs, is planned to further minimize the amount of sensitive user data being processed.
Additionally, the European Commission provides technical documentation, integration guides, and open-source components to support developers and online service providers. It includes hosted test services for demonstration and evaluation, detailed setup instructions, and a roadmap highlighting upcoming features. Ultimately, it aims to offer a consistent, secure, user-friendly age verification experience adaptable to different national contexts while promoting transparency and trust throughout the EU digital ecosystem.
This initiative bridges the gap until EU Digital Identity Wallets become widely available, ensuring that compliant and privacy-respecting age verification is accessible across digital services in the European Union.

It is, however, important to note that the EU's proposed representation of identity credentials is designed to comply with the mobile driving license (mDL) data standard specified in ISO/IEC 18013-5\cite{iso18013-5:2021}. While widely adopted, this approach became a subject of discussion, with cryptography experts arguing its insufficient security. This debate resulted in the publication of "Cryptographers' Feedback on EU Digital Identity's ARF"\cite{cryptographers_feedback_eudiw}, which stated that mDL lacks crucial privacy properties, such as unlinkability, required by eIDAS 2.0\cite{european_parliament_and_of_the_council_regulation_2024}.
The authors advocated adopting well-studied cryptographic primitives (e.g., Anonymous Credentials, BBS+ signatures) that address all the discussed issues. The drawbacks of these solutions, however, lie in their lack of standardization and the limited capabilities of the widely deployed hardware chips (e.g., mobile devices, smart cards). These limitations became a focus for further research to find an intermediate solution that could meet the security requirements while using the existing hardware. A notable resulting work by Frigo and Shelat\cite{anonymous_creds_from_ecdsa} was later acknowledged by the EU Commission, as a candidate to be standardized and implemented in the upcoming iterations of Technical Specifications for Age Verification\cite{UEAgeVerificationApp}.
As visible in recent publications\cite{poh_biometric_2025}, the search for secure age verification mechanisms is also evident in other domains of computer science (e.g., biometrics). However, there is a significant lack of solutions available for immediate implementation to address the needs and issues discussed above.

This context serves as the motivation for our proposal of an anonymous age verification service, which is discussed in detail in the next section. We present a solution based on an open-source, standardized Privacy Pass architecture and its anonymous tokens to enable privacy-preserving issuance and redemption of age credentials. 

\section{Solution - Anonymous Verification Age Tokens}
\label{sec:solution}

\begin{figure*}[!htp]
    \includegraphics[width=\textwidth]{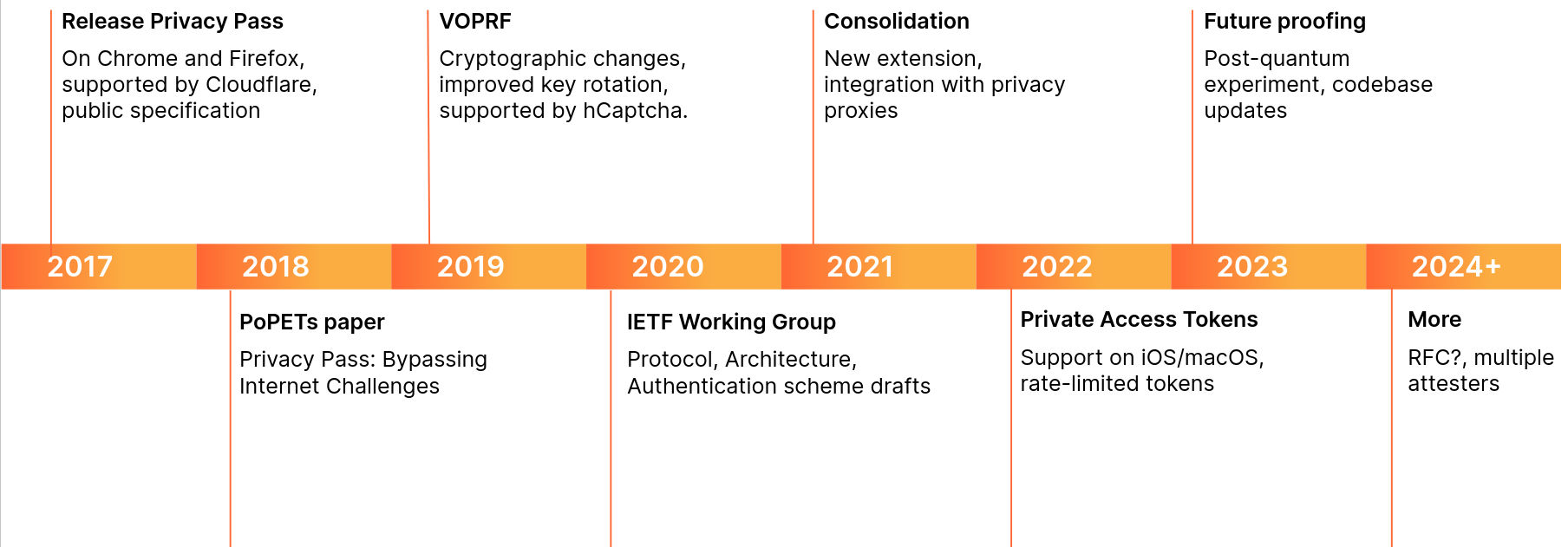}
    \caption{Privacy Pass Development Timeline by Cloudflare\cite{cloudflareBlog}} \label{fig:privacyPassTimeLine}
\end{figure*}

A Privacy Pass \cite{davidson_privacy_2018} token is a cryptographically signed, unlinkable digital credential issued to a user after they solve a challenge, such as a CAPTCHA. Users can then present these tokens to a service provider (e.g., a website) to prove their legitimacy without the service provider linking the token to the user's previous interactions, thereby improving privacy by enabling anonymous authentication and reducing the need for repeated challenges.
This architecture, proposed and maintained by Cloudflare (see Fig.~\ref{fig:privacyPassTimeLine}), reduced the number of CAPTCHAs presented to the honest web users by switching to other forms of user verification, including Privacy Pass tokens. This especially helped users who accessed online services via private networks (e.g., TOR or VPNs), since this group was most often queried for CAPTCHA challenges due to the very low reputation of their exit node addresses. The unlinkability property of these tokens enables privacy-conscious users to maintain their online anonymity without having to solve multiple inconvenient challenges.

Building on Cloudflare's implementation of Privacy Pass, Apple designed and deployed its own version of anonymous tokens, called the Private Access Tokens (PATs)\footnote{\url{https://developer.apple.com/news/?id=huqjyh7k}}. PATs work in tandem with the device's secure enclave, enabling attestation that users are not machines while binding the credential to the device.
In 2022, Apple devices began offering Privacy Pass tokens to websites that wanted to reduce CAPTCHA frequency by using the devices' hardware as an attestation factor. With this technology, service providers can request a PAT from the user before resorting to traditional visual CAPTCHAs. This native support does not require installing extensions or any user action to deliver a smoother, more private web browsing experience.

\begin{figure}[!htp]
    \includegraphics[width=\textwidth]{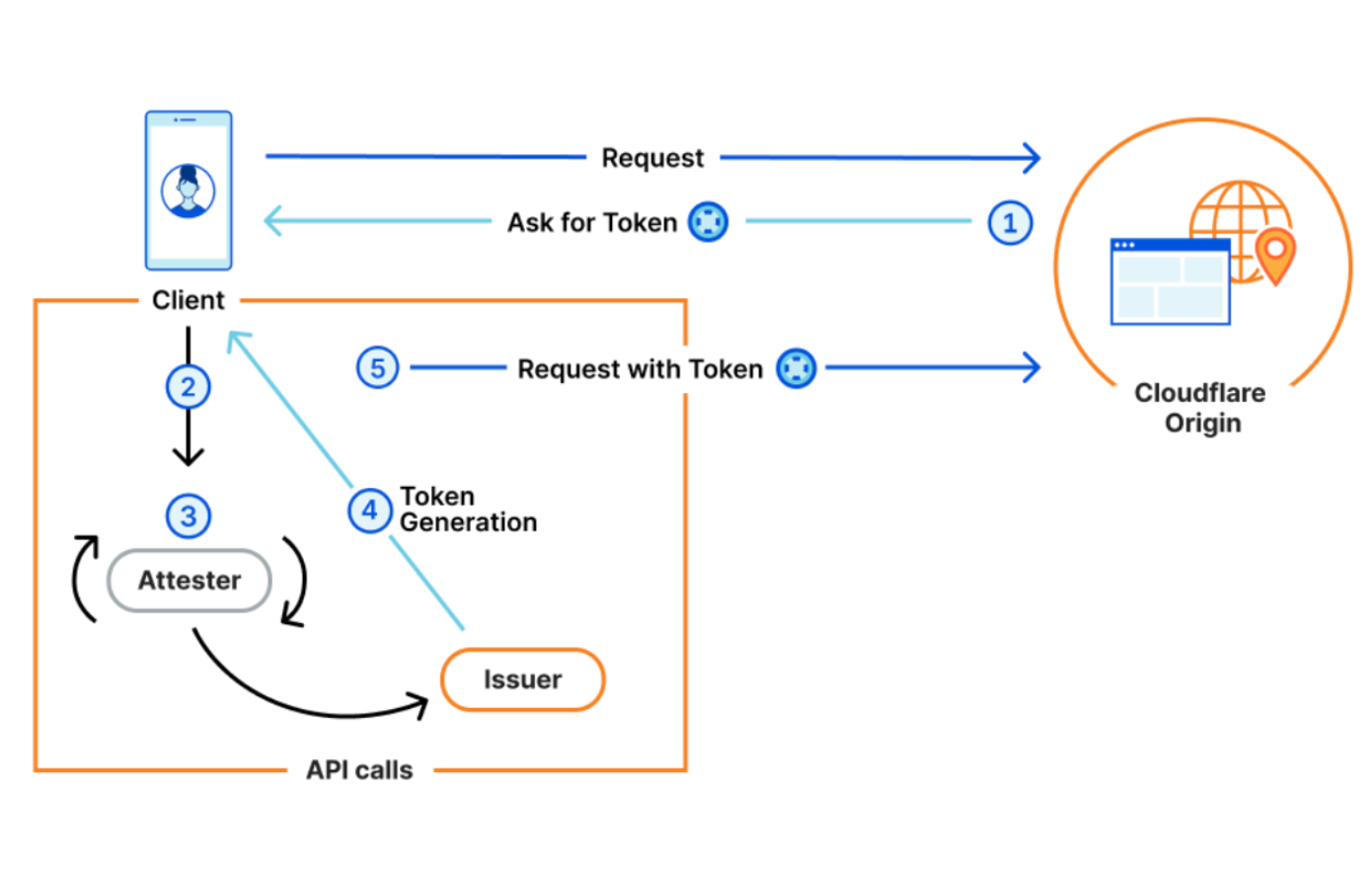}
    \caption{Privacy Pass Token Lifecycle by Cloudflare\cite{cloudflareBlog}} \label{fig:privacyPassTokenLifecycle}
\end{figure}

We focus on building our privacy-preserving age verification solution on the foundations established by Privacy Pass and Private Access Tokens technology. The main idea is to maintain the Privacy Pass architecture, depicted in Fig.~\ref{fig:privacyPassTokenLifecycle}, while substituting the CAPTCHA challenge presented to the user by the attester with a procedure for confirming the physical person's age. This technology benefits from narrowing the interaction between the physical user to just their mobile device (as in PATs) or the web browser extension (as in Privacy Pass web client\footnote{\url{https://privacypass.github.io/}}). This leaves the desired freedom to implement the final solution and integrate it with existing mechanisms that enable private companies on the market to electronically confirm users' age. Our proposition also achieves independence from an EU-based solution, allowing us to involve not only governmental but also private age attestation providers in the ecosystem, thereby reducing the risk of a single authority implementing a faulty solution.

\subsection{Sketch of our Proposal}

Our proposal incorporates the use of established, standardized technology, i.e., the Privacy Pass \cite{davidson_privacy_2018} architecture for the issuance and redemption of age verification attributes in the form of anonymous tokens. In the age verification scenario, tokens can be issued to the user who, instead of solving a CAPTCHA challenge, completes the Know Your Customer (KYC) procedure and meets specific age requirements. These tokens can then be spent during an interaction with a service provider serving age-restricted content as an anonymous proof that the user meets specific requirements, without revealing further information about their identity. 

The architecture of our solution is drawn directly from RFC 9576 \cite{davidson_privacy_2024}. The Privacy Pass system comprises two protocols and four parties: 
\begin{description}
    \item[client - ] user's device trying to get access to restricted content,
    \item[attester - ] party able to perform the KYC process, e.g., private company (for generic solution), Apple iCloud (for PATs),
    \item[issuer - ] certification authority, e.g., governmental institution,
    \item[origin - ] service providers hosting the restricted content.
\end{description}
The token issuance protocol enables the client to interact with the attester and issuer to receive age verification tokens.
These tokens are then used in the token redemption protocol, in which the origin grants access to age-restricted content upon the client providing a valid token, as depicted in Fig.~\ref{fig:privacyPassTokenLifecycle}.

When the client attempts to access restricted content guarded by the origin, they are requested to present an age verification token.
If the client already possesses one (e.g., a previously issued token), they can present it and receive access.
Otherwise, they start the token issuance protocol, interacting with the attester party, which serves as an institution capable of performing KYC procedures, ensuring the majority of the physical users.
Upon successful KYC, the attester forwards the client's token request to the issuer, which generates the anonymous token and provides it to the client.
This distinction between attester and issuer parties enhances client privacy, as the party providing the tokens receives no additional information about the client (e.g., data used for age verification purposes).
Furthermore, it enables the delegation of KYC and token-issuance procedures to independent institutions.

\begin{figure*}[!htp]
\includegraphics[width=\textwidth]{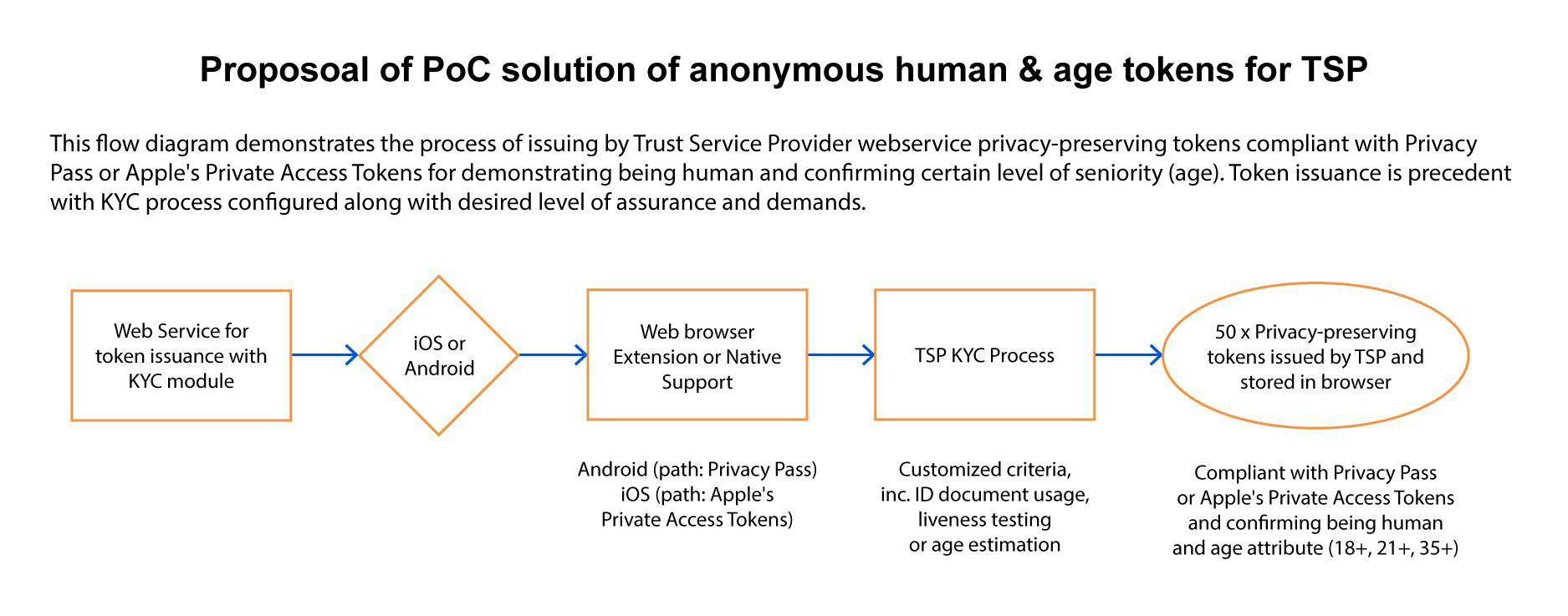}
\caption{Proposed Anonymous Age Verification Token Issuance Flow Diagram} 
\label{fig:solutionDataFlow}
\end{figure*}

Finally, the client can respond to the origin using the freshly generated token, thereby proving their majority.
When the origin successfully verifies the token, it may grant the client access to the safeguarded resources.
At the same time, to prevent double-spending of the same token within the origin's jurisdiction, the token is inserted to the revocation record, which is looked up for every token provided by clients.
Our solution emphasizes the use of publicly verifiable tokens, as described in RFC 9578 \cite{celi_privacy_2024}, based on the cryptographic primitive of blind signatures, standardized in RFC 9474 \cite{denis_blindrsa_2023}.
This allows any party with access to the issuer's public key to verify the token's validity, unlike privately verifiable tokens, where only the issuer (who possesses their own secret key) can verify the credential. Diagram of our solution is depicted in~Fig~\ref{fig:solutionDataFlow}.

\subsection{Properties and Extensions}

In this subsection, we informally introduce the properties of our system and discuss potential extensions and implementation approaches to achieve the strongest security guarantees.

\begin{description}
\item[Cryptographically guaranteed anonymity:] Publicly verifiable Privacy Pass tokens use a cryptographic primitive called a blind signature, which allows a user to obtain a valid, verifiable digital signature for a given message without revealing the message to the signer. In our case, the message and the signature under it form a single token. Privacy Pass uses the standardized blind RSA signatures \cite{denis_blindrsa_2023}, which provide the properties of blindness (i.e., the issuer does not learn the message during the signing protocol) and unforgeability (i.e., a malicious client cannot generate a valid signature without interacting with the issuer).
From the blindness property, follows the unlinkability of the tokens, meaning that the issuer is unable to match a signing transaction to an honestly signed token when presented with one.
This property also holds in the scenario of a collusion between the issuer and origins who received the token from the user. 

\item[Open and transparent solution:] The strength of the proposed architecture lies in the fact that the Privacy Pass technology is already standardized and well-established. 
Along with the RFC documents \cite{davidson_privacy_2024, pauly_privacy_2024, celi_privacy_2024}, the original work \cite{davidson_privacy_2018} came with an off-the-shelf, open-source blueprint implementations for the token issuing server\footnote{\url{https://github.com/privacypass/challenge-bypass-server}} and browser extension\footnote{\url{https://github.com/privacypass/challenge-bypass-extension}} serving as the client-side interface to the Privacy Pass ecosystem.
These codebases could serve as the foundation of the new solutions, such as the proposed age verification service.
The significant advantage of our approach is that the entire service can be instantiated as a web application, with a client-side browser extension handling token issuance and spending. Furthermore, for Apple devices browser extension is not mandatory, as the PAT ecosystem could be incorporated instead.
This significantly enhances the solution's accessibility by minimizing user-side setup requirements, eliminating the need for third-party software or additional mobile applications installed on the user's devices.
Regardless of the mobile platform the end user has access to, we propose a flowchart (see Fig.~\ref{fig:solutionDataFlow}) that enables the use of native Apple devices' environment of Private Access Tokens, as well as a solution based on a browser extension for Android users. Both are compatible with the Privacy Pass architecture and enable batch issuance of anonymous tokens.

\item[Double-Spending prevention:] A primary challenge in implementing anonymous, single-use tokens/credentials is the risk of token misuse across different origins.
Because the Privacy Pass architecture utilizes privacy-preserving mechanisms, such as blind RSA signatures \cite{denis_blindrsa_2023}, controlling the lifecycle of individual tokens becomes non-trivial.
This may allow a dishonest user to repeatedly attempt to access a protected resource using the same token.
To mitigate this, the Privacy Pass standard recommends two primary techniques: token context binding and tracking of the redeemed tokens.
In the first approach, the origin includes a chosen or randomly generated redemption context within the token challenge.
This context is then used by the client as part of a blindly issued token, ensuring its freshness and its restriction to a specific interaction with that origin.
However, this approach disqualifies the batch issuance of the tokens; therefore, we will not cover it further in this analysis.

The second approach allows the origin to detect double-spending locally by maintaining a record of spent tokens.
For every token presented, the origin checks a local database to determine if the token's unique nonce has already been used.
If a collision occurs, meaning that the client attempts to reuse the token dishonestly, the access is invalidated.
While straightforward to implement, this method faces scalability challenges when synchronizing records across different origins.
Without efficient synchronization, a dishonest user could potentially redeem a single token with multiple origins, which would not detect the double-spending.

\item[Batch token issuance:] To ensure that the Privacy Pass architecture remains seamless for the end user, the standard natively supports the batch issuance of the tokens.
This approach enables users to complete an attestation procedure (e.g., CAPTCHA, KYC-based age verification) and receive a bundle of cryptographic tokens in return.
Once successfully verified, the client can access multiple age-restricted resources until the token pool is exhausted.
This mechanism is particularly advantageous for age verification services, where the attestation process may be computationally expensive or require significant user engagement.
By enabling batch issuance, the architecture relieves users from the need for constant reverification, thereby improving usability while maintaining strict authorization requirements within defined policy restrictions, as reflected in specific age requirements in this scenario.
This separation of the issuance and redemption phases enhances privacy by making timing-correlation attacks significantly more difficult for potentially malicious parties, namely the issuer generating a token and the origin that requested a context-bound token.

\item[Solution scalability:] As demonstrated in the original work on PrivacyPass \cite{davidson_privacy_2018}, the batch approach to the anonymous token distribution scales effectively with the volume of issued tokens.
In this architecture, the anonymity of clients depends fundamentally on the size of the set of clients interacting with a specific issuer, known as the anonymity set.
If only a single client utilizes privacy-preserving tokens, any service provider can trivially deanonymize that user.
Our approach assumes a diverse ecosystem of multiple attesters and issuers, granting clients the autonomy to select service providers based on their service offerings, public reputation, and ease of use.
Status and identification data of entities offering attestation and token issuance, along with their cryptographic token verification keys, are to be stored in publicly available repositories, which aligns with the structure and governance model planned for trust service providers via the trusted lists across EU Member States \cite{european_parliament_and_of_the_council_regulation_2024}.
This approach simultaneously benefits from the ease of adding new origin parties to the ecosystem.
That is due to the fact that the token verification is local; it is supported only by a query to the public repository, which yields the cryptographic verification keys corresponding to the token issuer.

\item[Token issuer hiding:] While the Privacy Pass architecture is designed to protect user-oriented data, the standard redemption flow requires the client to reveal the issuer's name to the origin.
This disclosure allows the origin to identify which specific issuer generated a credential, potentially introducing privacy risks if that issuer serves a small anonymity set.
To mitigate the risk of surveillance and client linking based on their chosen token-issuing services, we propose incorporating an issuer-hiding technique.
In this scenario, issuers function as exchange points for tokens generated by other trusted entities.
If a client presents a valid age verification token, it can be spent to trigger the issuance of a new token, which the current issuer will validate.
This allows for mixing tokens held by a single user, significantly increasing the complexity of identity linking by dishonest relying parties who might otherwise use issuer-specific information to partition anonymity sets.
By decoupling the original issuance context from the final redemption context, this method ensures that the origin cannot effectively infer the original token provider.

\item[Device binding of tokens.]
Device binding is a security technique that restricts the use of cryptographic material to a specific physical device.
The ensurance that keys or tokens cannot be transferred or extracted from the device once generated, becomes essential in high-stakes environments where credential sharing or forgery carries significant consequences.
In the context of the anonymous age verification, the lack of device binding could lead to illegal distribution of honestly generated tokens.
This could expose minors to age-restricted content because the token's anonymity prevents the client's age from being verified during token redemption.
The current Privacy Pass architecture does not natively support direct device binding for individual tokens.
A possible mitigation of this problem may lie in adopting the challenge-response token issuance scheme, rather than the batch token issuance approach. By incorporating a timestamp as the token binding context, the token validity window could be adjusted and strictly limited, thereby hindering its potential for secondary distribution.
However, as previously noted, this transition to per-request issuance significantly increases service complexity and user engagement, as it may require clients to undergo a rigorous KYC age verification process for each access request.

\end{description}

\section{Conclusions}
\label{sec:conclusion}

The imperative to implement robust age verification mechanisms has gained unprecedented urgency, particularly in regions such as the United Kingdom, the European Union, and Australia, where regulatory frameworks are evolving rapidly; this trend is poised to extend globally in the coming years. Protecting minors from unrestricted exposure to harmful content remains paramount, as empirical evidence underscores the potential for such exposure to precipitate severe social, behavioral, and health disorders.

Contemporary deployed solutions have achieved only limited success, underscoring the need for comprehensive feedback mechanisms, rigorous lessons-learned analyses, and a fundamental redesign of deployment processes. A primary weakness lies in the overly broad scope of eligible methods for age verification and age estimation, which has engendered numerous vulnerabilities and attack vectors, culminating in documented instances of successful spoofing. Compounding this issue is the heterogeneous assurance levels across accepted methods, wherein some approaches exhibit markedly higher trustworthiness than others, thereby undermining system integrity.

Our analysis reveals that effective legislation must be inextricably linked to stringent technical standards; however, this integration carries the inherent risk of fostering monopolistic practices or over-reliance on government-sanctioned solutions. Key findings advocate for the substantial involvement of Trust Service Providers in age verification ecosystems to strengthen reliability and compliance. At the same time, regulators must avoid focusing exclusively on a single solution—as the saying goes, they must not “put all their eggs in one basket”—and instead promote a diversification of approaches to enhance resilience.

In summation, the market already offers viable technical means to engineer secure, privacy-preserving age verification systems. We have presented concept and explanatory architecture of such a framework using secure cryptographic components and existing open protocols.
Future efforts should prioritize standardized, high-assurance protocols, interdisciplinary collaboration, and iterative evaluation to safeguard digital spaces effectively.

\subsection{Remarks on Age Estimation Algorithms}
NIST launched a series of Face Analysis Technology Evaluation (FATE) projects, and one of them is dedicated to Age Estimation \& Verification\footnote{\url{https://pages.nist.gov/frvt/html/frvt_age_estimation.html}}.
The latest available NIST FATE AEV Evaluation Report (from May 2024)\cite{grother_face_2024} notes that age estimation accuracy has advanced since NIST's initial 2014 measurements. Still, for age estimation and metrics suited to "challenge" scenarios - where alternative age assurance methods are needed if the estimated age falls below a threshold - performance heavily depends on the algorithm, sex, image quality, region of birth, actual age, and interactions among these factors.

No single algorithm outperforms others across the board, with rankings shifting based on those variables. Accuracy also varies by metric, such as mean age estimation error versus false positive rates in challenge-25-style tests. Due to demographic variations, NIST expects developers to enhance capabilities over time. 

The NIST report presents results from the dataset employed in 2014 evaluation of age estimation algorithms. After a decade of research and development, coupled with major strides in deep neural network technology, five out of six algorithms now surpass the top-performing algorithm submitted to NIST in 2014. On a shared visa database, the best mean absolute error \textbf{(MAE)} has dropped from 4.3 to \textbf{3.1 years}.

These outcomes show that when using age estimation solutions based on facial images, we can still classify a 15-year-old as an adult or question the adulthood of a 21-year-old. This leads to the conclusion that this method should not be highly trusted and should not be used in sensitive scenarios such as access to restricted content.

\subsection{Limitation of the work}

While the proposed architecture leverages the standardized Privacy Pass protocol to provide a robust and privacy-preserving age verification service, several limitations remain that must be acknowledged:

\begin{description}
    \item[Token Device and Browser Binding:] The baseline of the Privacy Pass protocol standard cannot securely tie credentials to a specific user device. This transferability risk could allow users to share authorized tokens with unverified individuals. Addressing this through strict per-request issuance is technically possible, but it would severely degrade the user experience by eliminating the efficiency of batch verification, leading to significant user engagement and possible overload of the complex Know Your Customer procedure. The current solution limits the token binding to the client's honesty. Based on the assumptions underlying the EU's Age Verification solution, it is imperative that the user's device follows the protocol, which disallows token transfers between different devices, and thus binds the token to be spent only with the device it was issued to.

    \item[Double-Spending Prevention:] Preventing the reuse of anonymous, single-use tokens across multiple service providers presents a complex scalability challenge. Because the system relies on origins maintaining local records of spent token nonces, preventing cross-origin token reuse requires highly efficient, real-time synchronization across all participating service providers. Without such synchronization, a dishonest user could successfully redeem a single token at multiple independent origins before the double-spending is globally recognized.

    \item[Lack of Cryptographic Standardization:] The compromises present in this architecture fundamentally stem from the current landscape of cryptographic standardization. While advanced cryptographic primitives can natively resolve vulnerabilities related to device binding and double-spending without compromising user privacy, they currently lack both widespread standardization and native support on widely deployed consumer hardware chips. Until these advanced mechanisms achieve broader industry adoption, intermediate solutions based on established, standardized protocols like Privacy Pass remain the most viable, despite their inherent structural limitations.
\end{description}

%
%

\bibliographystyle{splncs04}
\bibliography{bibliography}

\end{document}